\documentclass[aps,twocolumn]{revtex4}

\input epsf
\usepackage{amsmath}
\usepackage{amsfonts}
\usepackage{amssymb}%

\begin{document}

\title{
Discovering the Higgs Through Highly-Displaced Vertices
}
\author{Matthew J. Strassler
and Kathryn M. Zurek}
\affiliation{Department of Physics,
P.O Box 351560, University of Washington,
Seattle, WA 98195\\
}
\begin{abstract}{
We suggest that the Higgs could be discovered at the Tevatron or the LHC
(perhaps at the LHCb detector) through decays with one or more substantially
displaced vertices from the decay of new neutral particles.  This
signal may occur with a small but measurable branching fraction in the
recently-described ``hidden valley'' models, hep-ph/0604261;
weakly-coupled models with multiple scalars, including those of
hep-ph/0511250, can also provide such signals, potentially with a much
larger branching fraction.  This decay channel may extend the Higgs
mass reach for the Tevatron.  Unusual combinations of $b$ jets, lepton
pairs and/or missing energy may accompany this signal.  }
\end{abstract}

\maketitle

If the Higgs boson is standard-model-like, then 
250--1000 Higgs production events (for $m_h$=200--120
GeV)  \cite{kilgore} have been produced in each Tevatron detector. 
Much recent work has focused on expanding the Higgs sector and the
possible decay channels of the Higgs boson
\cite{NMSSM,Wells,WeirdHiggs,Bargeretal}.  Most of
these results have negative implications for the Tevatron and LHC:
the signals they generate are as hard as or harder to
see than the standard ones. However, if a small but reasonable
fraction of Higgs boson decays are unusual and spectacular, this
special class of events may allow the Higgs to be found easily,
perhaps even with current data. 
In particular, we focus here on Higgs
decays to long-lived neutral particles that may decay at macroscopic
distances from the primary vertex.  

The purpose of this paper is to emphasize a simple but perhaps
underappreciated point.  The Higgs may serve as a window into a new
sector of particles uncharged under the standard model gauge group.
Its branching fraction to decay to these particles could be
substantial.  Some of these particles may have long lifetimes and may
decay visibly with a displaced vertex.  No fine-tuning is required to
allow this to occur.  The ease of building models with this
phenomenology, and the limited experimental constraints on light
neutral long-lived particles, argues that we should not be surprised
if the Higgs reveals itself first through such decays.  This, 
along with the fact that the new particles may be tied up with the
solution of the hierarchy problem, suggests
that searches for displaced vertices and associated
multi-jet/multi-lepton final states should be a high priority
for the Tevatron and the LHC.

Weakly-interacting models with possibly long-lived neutral particles
have arisen before; recent examples were given in \cite{WeirdHiggs}.
In \cite{hidval} a strongly-interacting ``hidden valley'' model,
inspired by top-down constructions such as those in
\cite{stringexamples}, was presented, where it was claimed that Higgs
mixing with another scalar field allows it to decay to two (or more)
new composite electrically-neutral particles (``v-hadrons'') which can
in some cases have long lifetimes.  Many other models have surely been
written down elsewhere.  It is easy to construct models where the
branching fraction to decays of this type is large enough to see in
current Tevatron data.  Even more models would allow discovery in this
channel toward the end of the Tevatron run or in the early days of the
LHC.  

Up to this point few Tevatron analyses have looked for 
long-lived neutral particles, excepting \cite{cdfZsearch,mumu}; no
general program of searches has been undertaken.  We are unaware
of any organized set of studies for the LHC.  We hope this
article will help motivate such a program, which would of course be
sensitive to a wide range of models well beyond those discussed below.

Such a program should consider many different final states.  The
lifetimes of the resonances are not constrained; decays at centimeter
and meter scales should equally be considered.  Each Higgs decay may
produce two or more resonances, with the multiplicity possibly varying
from event to event.  In many models, the resonances will decay to the
heaviest fermion pair available, with branching fractions similar to
those of the standard model Higgs.  For instance, this is the case in
some models of \cite{WeirdHiggs,hidval}.  Other resonances may decay
flavor-universally, with branching fractions to $\mu^+\mu^-$ and
$e^+e^-$ pairs that may be a few percent or more.  Decays to gluon
pairs are also possible.  For instance, a complex Higgs decay in a
hidden-valley model could produce, say, four resonances, one decaying
promptly to jets, one escaping the detector giving missing energy, and
two decaying to $b\bar b$, each with a displaced vertex.  But this is
only one of many possible decay modes in a typical model.
Clearly some final states are easier than others for triggering and
experimental analysis, but even a few displaced vertices might be the
key to finding these decays.  The branching fractions need not be
large for success; a few fully reconstructed decays, even with poor
resolution, would suffice for a claim of evidence of a new resonance
consistent with a Higgs boson.  The situation may be better still; the
branching fractions may be quite large for $m_h\ll 2m_W$.  As they may
remain measurable even for $m_h \agt 2m_W$, the Tevatron may be
considerably more sensitive to heavier Higgs bosons than is usually
thought.

Below we will confirm that it is reasonable for branching fractions to
neutral long-lived particles to be at observable levels.  We will
illustrate the simplicity of the model-building involved by briefly
considering a few cases.  We will start with two weakly-coupled
models, one nonsupersymmetric, one supersymmetric, with new
fundamental scalars.  Then we will discuss several hidden valley
models with new light composite particles.  After some general remarks
on more complex signatures, we will conclude with some comments
upon the implications for the ongoing and upcoming experiments.

\subsection{Two models with fundamental scalars}

To begin, we build a
simple scalar theory to illustrate how the Higgs can decay 
to two scalar resonances $X$, each of which decays
with a displaced vertex to heavy flavor.  
Consider adding a real scalar
$X$ to the standard model, and write the potential
\begin{eqnarray}\label{Vchi}
V &=& -\mu^2 H^2 + \lambda H^4+ M^2X^2 + \kappa X^4
+ \zeta X^2 H^2  \nonumber \\ 
& &+ aX  + b X^3+ c X H^2  \ ,
\end{eqnarray}
where $H^2\equiv H^\dagger H$.
We assume here that TeV-scale physics protects the
masses and dimensionful couplings, and that, after electroweak
symmetry breaking (EWSB), $m_h>2 m_X$, so that $h\to XX$ can oocur.

When $a=b=c=0$, the theory has a ${\bf Z}_2$ symmetry under $X\to-X$.
If $\langle X\rangle=0$, so this symmetry is not spontaneously broken,
then $X$ is stable, and the decay $h\to XX$ will be invisible.  For a
light Higgs, its branching fraction $\sim |\zeta v /y_b m_h|^2$
need not be small; here $v=\sqrt 2\langle H\rangle$ and $y_b$ is 
the $b$-quark Yukawa coupling.  If instead $a,b,c$ are
nonzero but small, the ${\bf Z}_2$
symmetry is explicitly but softly broken.
It is then technically natural to have small mixing, after
EWSB, between $X$ and $H$.  While the mixing does
not much affect $H$ decays, it drastically affects $X$ decays (since
without it $X$ is stable.)  If the low-mass eigenstate is of
the form $|X\rangle + \epsilon |h\rangle$, the decay rate for $X$ is
$\Gamma_X = {\epsilon^2}\Gamma_h$, and its branching fractions are
those of the Higgs.  Since $\epsilon$ is related to the breaking of
the ${\bf Z}_2$ symmetry, it can naturally be small; the decays may be
prompt, invisibly long, or anything between.

If $\epsilon$ were not small, and the $X$ decays were prompt, the
above model would have similar phenomenology to the next-to-minimal
supersymmetric standard model (NMSSM) in certain regimes
\cite{NMSSM,WeirdHiggs} where the Higgs branching fraction to light
neutral states can be large.  We have not attempted to modify the
NMSSM itself to obtain long-lived states (but see \cite{WeirdHiggs}.)
Long-lived states in Higgs decays can also be obtained in a
supersymmetric variant of the model in Eq.~(\ref{Vchi}). Consider the
MSSM with electroweak singlet scalars $S$ and $X$, and superpotential
\begin{equation}\label{SXX}
W = \lambda S H_uH_d + \eta S^3 + \kappa S X X + b X^3
\end{equation}
If $b=0$, this model has an $X\to -X$ symmetry.  After EWSB and
supersymmetry (SUSY) breaking, $S$, $H_u$ and $H_d$ acquire expectation
values.  We assume $m_h>2 m_X$.  We may (naively) require
$\lambda\kappa\alt 1/20$ to avoid a large $X$ mass; even so, since
$y_b\ll 1$, we can have $Br(h\to X X)\propto |\lambda \kappa
v/y_bm_h|^2$ of order one.  If $b\neq 0$, then after SUSY breaking a
loop effect will allow $X$ to mix with $S$, and thereby with $h$, with
a mixing angle $\epsilon\alt |\kappa b|/16\pi^2$.  Then $X$ can decay
to heavy flavor with a long lifetime.  The angle $\epsilon$ may be further
suppressed if the mass splitting between $X$ and its fermion partner
$\psi_X$ is rather small, as is the case in some SUSY-breaking
scenarios.  Thus there is no obstruction to building natural models
with a Higgs that often decays in this fashion. 

If $Br(h\to X X)\agt .1$, displaced $\tau$ pairs may be detectable;
they are lightly constrained by \cite{mumu}.  If in addition $m_X<
2m_b$, so that $Br(X\to\tau^+\tau^-)\sim 1$, then \cite{mumu} is a
more significant constraint.  More complex decays, analogous to those
in \cite{WeirdHiggs}, can also potentially occur in this model; we
will briefly return to this later.


For a Higgs with $m_h\agt 2m_W$, $Br(h\to X X)\propto |\lambda \kappa
v^2/m_h^2|^2$ is perhaps of order one percent or so; also $\Gamma_h$
is larger now, so a smaller $\epsilon$ is needed for displaced
vertices to result.  If this were the case, the reach of the
Tevatron would extend toward much higher $m_h$ than is normally
considered.  Despite a smaller rate and trigger efficiency, events
with displaced vertices may well be more important than $WW$ decays to
leptons, for which there is an irreducible background and no
possibility of kinematic reconstruction.  Also, in two-Higgs doublet
models there is a CP-odd scalar $A^0$  which is produced in $gg\to A^0$ but
which cannot decay to $WW$ or $ZZ$.  Its discovery may be
made much easier by exotic decays with a
large branching fraction.

\subsection{Composite resonances in hidden-valley models}

We now turn to a different class of models.  The phenomenology of
confining hidden valley models was recently outlined in \cite{hidval}.
These models can show qualitatively similar signals to the theories
just discussed, though the origin of the signals is quite different.
We will now see that the illustrative models of \cite{hidval} can give
the Higgs a substantial branching fraction to long-lived neutral 
resonances.

\begin{figure}[htbp]
  \begin{center}
    \leavevmode
     \epsfxsize=.36 \textwidth
     \hskip 0in \epsfbox{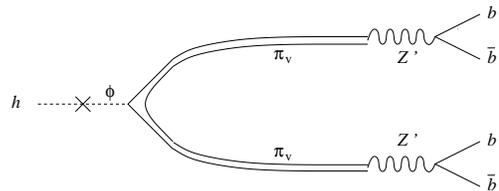}
  \end{center}
\caption{Higgs decay to v-hadrons, each of which
decays to $b\bar b$.
}
\label{fig:hdecay}
\end{figure}

We briefly summarize the particular hidden-valley models that were
explored in \cite{hidval}.  (Hidden valleys --- sectors with a
non-abelian gauge group under which no standard model matter is
charged, which couple weakly to the standard model via higher
dimension operators, and which have a mass gap --- are common in
string constructions of the standard model \cite{stringexamples},
though string theory is of course not required.)  The ``v-sector''
consists of a confining gauge group that makes v-hadrons out of its
v-quarks, in analogy with QCD.  The only couplings between the
v-sector and the standard model occur through a heavy $Z'$ (which has
coupling $g'$ and mass $m_{Z'}=2\sqrt 2 g' \langle\phi\rangle$, where
$\phi$ is a scalar that also gives mass to v-quarks) and through
possible mixing (as in \cite{Wells}) between $\phi$ and the Higgs.  If
such mixing is present, a Higgs decay to v-hadrons, shown in
Fig.~\ref{fig:hdecay}, can be followed by the late decay of the
v-hadrons via the heavy $Z'$.
LEP I constraints conservatively require
$\hat v \equiv \sqrt 2 \langle\phi\rangle>5$ TeV, 
though this can often be relaxed,
perhaps to 2--3 TeV, because of conservation laws that render
some $Z$ decays invisible or forbidden.
In these simple models, lifetimes for
light pseudoscalars (which decay to heavy flavor) of mass
$\sim 20$ (40) GeV are of order 100 ps (1 ps); this includes the
$\pi_v^0$ [$\eta'_v$] for two [one] light v-flavors.  However, other
v-hadrons could give displaced vertices if an approximately-conserved
quantum number delays their decays. Additional
details are given in \cite{hidval}.

The potential for the scalar fields takes the form
\begin{equation}\label{V1}
V = -\mu^2 |H|^2 - \hat \mu^2 |\phi|^2
+\lambda |H|^4 + \tau |\phi|^4
+\zeta  |\phi|^2|H|^2
\end{equation}
Since 
$\gamma\equiv v/\hat v\sim 1/10-1/20$, we can set
$\zeta\sim\lambda\gamma^2$, $\tau\sim \lambda\gamma^2$, and 
then obtain a mixing angle $\theta$ which is naturally 
of order $0.1$ or larger.
Extreme
fine-tuning of parameters (at the classical level) is not required. 
Adjusting $\lambda$ and $\mu$, we can obtain acceptable
values for $m_h$ and $v$.


Given that the mixing angle is not too small, let us now estimate the
branching fraction.  The produced Higgs state is $\cos\theta|h\rangle
+ \sin\theta|\phi\rangle$, with $\sin\theta\alt .3$.  The production
process is slightly suppressed, by $\cos^2\theta$. 
Let $y_0$ be
the Yukawa coupling of $\phi Q\bar Q$ where $Q$ is the heaviest
v-quark allowed kinematically; then (assuming only minor phase space
suppression in the $h\to Q\bar Q$ decay) the branching fraction to
v-hadrons for a Higgs below 140 GeV is naively of order
\begin{equation}
\label{Brtovhads}
{y_0^2\over y_b^2}\sin^2\theta 
\approx \left[ {m_Q\over m_b} 
{v\over \hat v}\sin\theta \right]^2
\end{equation}
As a figure of merit, requiring
a branching fraction of 1 percent would require $\sin\theta\agt m_b/m_Q$.
However, there are large v-hadronic corrections to this result
\cite{HHG}.  We will not discuss this at length, but suffice it to say
that for the exclusive channel $h\to\pi^0_v\pi^0_v$ a conservative
estimate is given by replacing $m_Q$ with $m_\pi^2/3 m_h$ in
(\ref{Brtovhads}); this is a considerable underestimate in models
with additional heavy v-quarks \cite{HHG}.  If other v-hadron decay
channels are available, they may be favored over $\pi_v\pi_v$; some of these
may give displaced vertices as well \cite{hidval}.

The v-hadronic branching fraction of the Higgs may be larger in a
model with two $\phi$ fields (with expectation values $\hat
v_1$ and $\hat v_2$) and a single Higgs doublet $H$.
For instance, if $\tan\beta_v\equiv \hat v_2/\hat v_1\sim 20$, so that
$\hat v_1\sim v$, and if $\phi_1$ couples to $Q$, then the factor
$v/\hat v$ in (\ref{Brtovhads}) is enhanced by $\tan\beta_v$.  The
$\pi_v^0$ decay still proceeds through the $Z'$ propagator, and its
lifetime remains long.  However, if the standard model
sector also has two Higgs doublets
$H_1,H_2$, as in any
supersymmetric model, then mixing between the CP-odd scalars in the
two sectors can allow the $\pi^0_v$ to decay more rapidly,
eliminating the displaced vertices.  This can be
avoided only if $m_{\pi_v}<2m_b$, or if the standard model sector's
CP-odd scalar $A^0$ is heavier than a TeV or more, suppressing its
mixing angle relative to $\theta$ by $\sim m_h^2/m_A^2$.  However we
emphasize that this constraint is special to the simplest models.

In fact it is instructive to consider 
a simple supersymmetric generalization of
the hidden-valley model in \cite{hidval}.
Minimally supersymmetrizing the models of
\cite{hidval} requires two Higgs multiplets $H_u$ and $H_d$ and two
new v-Higgs multiplets $\phi_1$ and $\phi_2$.  The D-term of
the new $U(1)$ induces mixing between the
Higgses in this model, but generally is either too small or generates
a large shift in $m_h$ that must be cancelled elsewhere.
An F-term is therefore necessary to induce larger mixing.
If we add a singlet $S$ which couples
through the superpotential
\begin{equation}
W = \lambda SH_uH_d + \eta S^3 + \kappa S\phi_1\phi_2
\end{equation}
then it is possible to arrange for substantial $h-\phi$ mixing.
However, to keep the $\pi_v^0$ long-lived requires the CP-odd scalar
$A$ to be heavy or its mixing angle to be small.  A small mixing angle
can still allow branching fractions large enough to see at the LHC,
but to reach branching fractions of one percent while still retaining
visibly-displaced $\pi_v^0$ decays requires uncomfortable tuning.
Nevertheless, one should not conclude that supersymmetric
hidden-valley models disfavor decays of this type at the Tevatron.
SUSY also may introduce flavor-changing neutral currents
through off-diagonal v-squark mass-squared terms or A-terms, whose size depends
on the details of supersymmetry breaking.  Such terms could
allow an otherwise stable v-hadron,
such as the $\pi^+_v$ in the model of \cite{hidval}, 
to decay to $b\bar b$ with a long
lifetime \cite{hidval},
even if the $A$ is light.  (This can occur in the model of \cite{hidval}
for $q_+=-2$.)   In other words, the
$\pi^+_v$ lifetime is determined through the same mechanism used in
the weakly-coupled models above: a weakly-broken approximate symmetry.
In turn, this relaxes the constraint on $m_A$ and its mixing angle,
and therefore can allow $Br(h\to \pi^+_v\pi^-_v)$ to be much larger.
The increased number of events widens the range of $\pi_v^+$ lifetimes
for which a few displaced vertices could be observed now or in the near
future.

Of course this is just one of many hidden-valley models.  It provides
an existence proof that the Tevatron may be able to observe the Higgs
in this decay mode, and should be sufficient to motivate an
experimental search for $h\to (b\bar b)(b\bar b)$ with displaced vertices.
Tau pairs should also be sought.
But we must be prepared for other phenomenology as well \cite{hidval}.
A systematic classification of the most promising final states is
needed, though it will inevitably suffer from considerable
model-dependence.  Here we limit ourselves to some general remarks.


\subsection{More complex final states}

In weakly coupled models, including those of \cite{WeirdHiggs} and
that of Eq.~(\ref{SXX}), a more complicated spectrum will allow a
cascade of immediate decays, leading to a final state with several
long-lived particles.  In the hidden valley models, the Higgs will
preferentially decay to the heaviest kinematically allowed v-quarks.
These in turn form v-hadrons which may then decay to multiple
v-hadrons made from lighter v-quarks; or, if the confinement scale is
sufficiently low compared to $m_h$, showering and hadronization may
occur, resulting in multi-v-hadron final states \cite{hidval}.  (In
such a scenario, where the production rate is increased by heavier
v-quark masses, but the final state has a a large multiplicity of
v-hadrons whose decay rates are decreased by lighter v-quark masses,
the probability of seeing displaced vertices is enhanced.)  The
multiple resonances produced in these Higgs decays may decay with a
variety of lifetimes.  Not all the resonances may decay preferentially
to heavy flavor, and some may have complex decays.  Some unusual
possibilities are discussed briefly in \cite{hidval}.


If many or most Higgs decays of this type contain invisible particles
or have resonances decaying to soft particles, reconstructing the
Higgs resonance using these events may be very challenging.  Even so, the
kinematics of a large sample of events may allow a rough estimate of
the mass scale involved and hint that the initial state is $gg$.  At
the other extreme, if all the new particles decay visibly but
promptly, as in \cite{NMSSM, WeirdHiggs}, then even selecting events
may be extremely difficult.  This unfortunate outcome cannot be
excluded, but neither is there any reason to expect the worst.

\subsection{Implications}

One could explore many more examples, but this would merely bolster a
point that should already be clear.  With possibly hundreds 
of Higgs boson events on tape, various searches for
displaced vertices in Tevatron data could be undertaken now.
Meanwhile, it seems to us that the LHC experiments should prioritize
ensuring that tracking software is efficient at detecting displaced
vertices, especially those involving muons and/or multiple charged
tracks.  Both CMS and ATLAS will implement a displaced-vertex trigger
(often called a ``$b$-trigger'') as part of their high-level triggers.
We would suggest that the upper-limit on the vertex displacement be
extended as far as possible and not be constrained by expectations
from $b$ decays.  It may also be valuable to consider special-purpose
triggers for (1) highly-displaced many-track vertices including a 
muon or electron, (2) hard
calorimeter jets with no stiff tracks pointing from the pixel detector,
and/or (3) calorimeter jets with no electromagnetic energy or
associated stiff tracks.  To the extent detector effects and QCD
backgrounds make it difficult to use such objects in triggering, one
could consider requiring two such objects simultaneously.  This should
still be efficient, since two or more displaced vertices per event
will be common in many models.  In general, since it appears that
light, long-lived neutral resonances (without charged resonances) are
a common feature of reasonable models, there are good reasons to
maximize the sensitivity of the LHC experiments to their
phenomenology.  Indeed it is possible that the LHCb experiment is ideally
suited for detecting and studying such states.

We have argued that it is possible in a diverse array of models for
the Higgs to decay to two or more neutral resonances, with a wide
range of possible final states. Some of these resonances
may decay
with a displaced vertex anywhere in the detector.  
The branching fractions for these decays may allow them to be
discovered at the Tevatron with current or future data, and easily
found at the LHC; this is especially true for a light Higgs but may
even be so for $m_h\agt 2m_W$.  Moreover, signs of other new
phenomena, such as those of \cite{hidval}, or perhaps supersymmetry
itself \cite{susyvsector}, could be discovered first in such searches.
We urge the Tevatron and LHC communities to take this possibility
seriously and undertake a wide array of searches for displaced
vertices.  A comprehensive program is needed, not only because of the
many possibilities offered by different models, but also because, even
within one model, the Higgs may have many decay modes.  It may be that
only in the combination of multiple analyses can the Higgs boson be
discovered in the immediate future.  Although many of these analyses
will be technically difficult and time-consuming, it would seem that
the possibility of discovering the Higgs more rapidly than typically
expected provides sufficient motivation.

We thank A. Haas,  H. Lubatti, A. Nelson, S. Stone
and N. Weiner for discussions.
This work was supported by U.S. Department
of Energy grants DE-FG03-00ER41132 (KZ) and DE-FG02-96ER40956 (MJS).

\end{document}